\documentclass[twocolumn,pre,superscriptaddress]{revtex4}
\usepackage{latexsym}
\usepackage{graphicx}
\begin{document}

\title{Flow properties of driven-diffusive lattice gases: theory and computer simulation}

\author{Debashish Chowdhury}
\affiliation{Department of Physics, Indian Institute of Technology,
Kanpur 208016, India.}
\author{Jian-Sheng Wang}
\affiliation{Department of Computational Science,
National University of Singapore, Singapore 119260,
Republic of Singapore.}

\date{9 June, revised 14 August, 2001}

\begin{abstract}
We develop $n$-cluster mean-field theories ($1 \leq n \leq 4$) 
for calculating the {\it flux} and the {\it gap distribution} 
in the non-equilibrium steady-states of the Katz-Lebowitz-Spohn 
model of the driven diffusive lattice gas, with attractive and 
repulsive inter-particle interactions, in both one and two 
dimensions for arbitrary particle densities, temperature as well 
as the driving field. We compare our theoretical results with the 
corresponding numerical data we have obtained from the computer 
simulations to demonstrate the level of accuracy of our 
theoretical predictions. We also compare our results with those 
for some other prototype models, notably particle-hopping models 
of vehicular traffic, to demonstrate the novel qualitative 
features we have observed in the Katz-Lebowitz-Spohn model, 
emphasizing, in particular, the consequences of {\it repulsive} 
inter-particle interactions.
\end{abstract}

\pacs{PACS number(s): 05.60.-k, -89.40.+k}

\maketitle

\section{Introduction}

The driven-diffusive lattice gas models are of current interest 
in non-equilibrium statistical mechanics \cite{sz,gs,privman,md}. 
Depending on the nature of the drive, these driven-dissipative 
systems can attain steady-states that are far from equilibrium. 
The simplest driven-diffusive lattice gas model that incorporates 
inter-particle interactions is the Katz-Lebowitz-Spohn model 
\cite{kls} (from now onwards referred to as {\bf KLS}). Some of 
the particle-hopping models of vehicular traffic \cite{css} are 
closely related to some special limits of the KLS model in one 
dimension. Therefore, in order to compare and contrast the 
spatio-temporal organizations and the flow properties of the KLS 
model with those in the particle-hopping models of vehicular 
traffic, we calculate here those properties of the KLS model 
which  are important from the perspective of vehicular traffic.  

Over the last decade extensive investigations of vehicular traffic 
have been made using the so-called particle-hopping models 
which represent each vehicle by a particle \cite{css,helbing,tgf}. 
All these traffic models are defined on discrete lattices each 
site of which, in the spirit of the lattice gas models, represents 
a cell that can accommodate at most one particle at a time. 
In almost all the standard particle-hopping models of vehicular 
traffic the only non-vanishing inter-particle interaction is 
the mutual hard-core repulsion which is usually implemented through 
the condition of exclusion principle: no two particles are allowed 
to occupy the same lattice site simultaneously. Therefore, a 
comparison of our results on the KLS model with the corresponding 
results for the particle-hopping models of vehicular traffic will 
show the effects of inter-particle interactions other than mere 
hard-core repulsion. 

The {\bf flux} (per lane) is defined to be the number of particles 
(per lane) crossing a detector site per unit time. In the context 
of vehicular traffic \cite{may}, the most important quantity of 
interest is the so-called {\it fundamental relation} which depicts 
the dependence of the flux on the density of the vehicles. The 
number of empty sites in between a pair of particles is usually 
taken as a measure of the corresponding distance-headway {\bf DH}. 

In this paper we theoretically calculate the DH distributions  
and the flux in the steady-states of the KLS model, separately 
for attractive and repulsive inter-particle interactions, 
within the framework of a {\it cluster mean-field theory} ({\bf MFT}) 
\cite{md,dickman86,dickman,attila91,attila93,attila96} which 
has been very successful also in the theoretical treatment of the 
particle-hopping models of vehicular traffic \cite{css,ssni,chowdh,chowth}. 
We also indicate the level of the accuracy of our cluster-MFT 
results by comparing these with the corresponding numerical data 
obtained from our computer simulations of the KLS model. 

The organization of this paper is as follows: in section II we 
define the KLS model and some related particle-hopping models 
which are relevant for our discussion in the subsequent sections. 
We summarize in section III the methods of the cluster MFT we 
use for our theoretical calculations as well as those of computer 
simulation. In sections IV, V and VI, we present our theoretical 
results for the one-dimensional KLS model (both with attractive 
and repulsive interactions) in the $1$-cluster, $2$-cluster and 
$4$-cluster approximations, respectively, together with the 
corresponding numerical data from our computer simulations. We 
present our results for the two-dimensional KLS model in section 
VII. We compare and contrast the results for the KLS model with 
the corresponding results for the particle-hopping models of 
vehicular traffic in section VIII before summarizing the main 
results in the concluding section IX.

\section{The models}

\subsection{The KLS model}

Suppose the variable $c_i$ describes the state of occupation of 
the site $i$ ($i=1,2,...,N$) on a discrete lattice; $c_i$ is 
allowed to take one of the only two values, namely, 
$c_i = 1$ if the site $i$ is occupied by a particle and $c_i = 0$ 
if it is empty (or, equivalently, occupied by a ``hole"). The 
Hamiltonian for the system, in the absence of any external driving 
field, is given by 
\begin{equation} 
{\cal H} = - 4 J \sum_{<ij>} c_i c_j, 
\label{eq-1}
\end{equation}
where the summation on the right hand side is to be carried out 
over all the nearest-neighbor pairs and $J$ takes into account 
the corresponding inter-particle interactions.

The KLS model can be recast in the language commonly used in the 
theory of magnetism by using classical Ising spin variables 
$S_i = (2c_i - 1)$ where $S_i = 1$ and $S_i = -1$ represent the 
particles and holes, respectively, and the corresponding 
Hamiltonian, in the absence of the external drive, is given by 
\begin{equation}
{\cal H}' = - J \sum_{<ij>} S_i S_j. 
\label{eq-2}
\end{equation}
The {\it attractive} and {\it repulsive} inter-particle interactions, 
captured by $J > 0$ and $J < 0$, respectively, in the Hamiltonian 
(\ref{eq-1}) correspond to the ferromagnetic and antiferromagnetic 
interactions in the form (\ref{eq-2}) of the Hamiltonian. 

However, throughout the rest of this paper, we shall use the 
particle-hole picture, where the instantaneous state (configuration) 
of the system at time $t$ is completely described by $(\{c\};t)$. 
For example, in case of a system of length $L$ in dimension $d=1$, 
$(\{c\};t) \equiv (c_1,c_2,...,c_L;t)$. 
Similarly, for the $L_x \times L_y$ square lattice
$(\{c\};t) \equiv (c_{11},c_{12},..,c_{ij},...,c_{L_xL_y};t)$. 
The average density $c$ of the particles is given by 
$c = \lim_{N \rightarrow \infty, L \rightarrow \infty} N/N_s 
= \lim_{N \rightarrow \infty, L \rightarrow \infty} (\sum_{i}^{N_s} c_i)/N_s$ 
where $N_s$, the total number of available sites is $L$ for a 
linear chain and $L^2$ for a square lattice of size $L \times L$. 
Note that, because of the conservation of the particles, the 
density $c$ is conserved by the dynamics.

The dynamics of the system is governed by the well-known Kawasaki 
dynamics: at any non-zero temperature $T$, a randomly 
chosen nearest-neighbor particle-hole pair is exchanged with the 
probability $\min[1,e^{-\beta(\Delta {\cal H} + \ell E)}]$ where 
$\beta = (k_BT)^{-1}$ ($k_B$ being the Boltzmann constant) and 
$\Delta {\cal H} = {\cal H}(\{c\}_{new}) - {\cal H}(\{c\}_{old})$ 
is the difference in the energy of the new and old configurations 
while $\ell = (-1, 0, +1)$ for jumps, respectively, along, 
transverse to, against the direction of the driving field $\vec E$.
Throughout this paper we take $k_B = 1$ and express the temperature 
$T$ in the units of $J$.

For the KLS model with {\it attractive} inter-particle interactions 
($J > 0$) on a square lattice, there is not only an ordered state  
at all $T < T_c(E)$, but the critical temperature $T_c(E)$ 
increases with $E$, saturating at a value 
$T_c(E \rightarrow \infty) \simeq 1.4 T_c(E=0)$ 
where $T_c(E=0)$ is the critical temperature of the corresponding 
Ising model in thermodynamic equilibrium \cite{sz,kls}. On the other 
hand, $T_c(E)$ decreases with $E$ when the inter-particle interactions 
are {\it repulsive} (i.e., $J < 0$); 
the ordering is altogether destroyed by sufficiently large $E$.
However, there is no ordered structure at any non-zero temperature 
in the one-dimensional KLS model, irrespective of the sign of 
the interaction $J$. Because of this intrinsic qualitative 
difference in the nature of the ordering in the steady states in 
$d = 1$ and $d = 2$ we present the corresponding results in  
separate sections.

\subsection{The Totally Asymmetric Simple Exclusion Process} 

In the Totally Asymmetric Simple Exclusion Process (from now 
onwards referred to as {\bf TASEP}) \cite{spohn}, initially, 
$N$ classical particles occupy {\it randomly} the sites of a 
one-dimensional lattice of length $L$ ($\geq N$). One time 
step of the dynamics consists of updating the position of $N$ 
particles picked up in a random-sequential manner; each randomly 
chosen particle moves forward, with probability $q$, if the 
lattice site immediately in front of it is empty. For this 
model, the simple single-site (i.e., $1$-cluster) MFT 
\cite{derrida} gives the {\it exact} flux 
\begin{equation}
F = q c (1-c) 
\label{eq-3}
\end{equation}
for all densities $c$ of the particles.

\subsection{Comparison between the models}

In the special case $E = 0$ the KLS model reduces to the corresponding 
standard Ising model in thermodynamic equilibrium. Note that in 
$d = 1$, in the opposite limit $E \rightarrow \infty$ the hopping 
against the driving field becomes impossible and, moreover, the field 
$E$ dominates so overwhelmingly over $\Delta {\cal H}$ at all non-zero 
temperatures $T$ that, in this limit, the one-dimensional KLS model 
reduces to the TASEP, with $q = 1$, irrespective of the sign of the 
interaction $J$, provided $J$ remains {\it finite}. However, in $d=2$, 
in the limit $E \rightarrow \infty$ the hopping probabilities in the 
directions transverse to the driving field remain finite and the ratio 
of the hopping probabilities in the direction of the field and those 
in the transverse direction diverges \cite{beiren}. For all the 
non-vanishing finite $E$, in the limit $J = 0$, the one-dimensional 
KLS model reduces to the TASEP with the hopping probability 
$q = \min\{1,e^{-\ell E/k_BT}\}$. Finally, at infinitely high 
temperatures each particle moves completely randomly, independent of 
each other, with equal probability in all directions.

\section{methods of calculation}

In this section we briefly outline the methods 
of our analytical as well as numerical calculations.

\subsection{Cluster-mean-field theory}

The dynamical cluster MFT has been used successfully in the 
analytical treatments of several non-equilibrium models including, for 
example, surface-reaction models \cite{dickman86} and particle-hopping 
models of vehicular traffic \cite{ssni,chowdh,chowth}. However, in all 
the traffic models there is no inter-particle interaction except, of 
course, the hard-core repulsion. Moreover, unlike the traffic models, 
the particles in the KLS model can also move against the drive. 

In this paper we extend the approach in appropriate manner to 
calculate the flux $F$ as a function of $c$ in the KLS model for 
arbitrary $J \neq 0$, $0 \leq T \leq \infty$ and $\vec E \neq 0$.

We define a $n$-cluster ($n < N$) to be a collection of $n$ sites 
each of which is the nearest-neighbor of at least another site 
belonging to the same cluster. For simplicity of notation, let us 
consider $d = 1$. We denote the probability of finding an $n$-cluster 
in the state $(c_1,c_2,...,c_n)$ at time $t$ by the symbol 
$P_n(c_1,c_2,...,c_n;t)$. We treat an $n$-cluster exactly and 
approximate all the $n+m$-cluster probabilities by a product of 
$n$-cluster probabilities in a manner so as to couple the 
$n$-cluster to the rest of the system  self-consistently (see, 
for example, \cite{css} for a pedagogical introduction and the 
existing literature). 

It is straightforward to see that the state of the $2$-cluster 
$(c_i,c_{i+1})$ at time $t+\Delta t$ depends on the state of 
the $4$-cluster $(\tau_{i-1},\tau_i,\tau_{i+1},\tau_{i+2})$ at 
time $t$ so that the exact master equation 
\begin{eqnarray}
&\biggl[&\frac{dP_2(c_i,c_{i+1})}{dt}\biggr] \nonumber \\
&=& \sum_{\tau_{i-1},\tau_{i+2}} \biggl[P_4(\tau_{i-1},c_{i+1},c_i,\tau_{i+2}) 
w(\tau_{i-1},c_{i+1},c_i,\tau_{i+2}) \nonumber \\
&-&P_4(\tau_{i-1},c_i,c_{i+1},\tau_{i+2}) 
w(\tau_{i-1},c_i,c_{i+1},\tau_{i+2})\biggr] \nonumber \\ 
&+& \sum_{\tau_{i-1},\tau_{i-2}} \biggl[P_4(\tau_{i-2},c_i,\tau_{i-1},c_{i+1}) 
w(\tau_{i-2},c_i,\tau_{i-1},c_{i+1}) \nonumber \\ 
&-& P_4(\tau_{i-2},\tau_{i-1},c_i,c_{i+1})   
w(\tau_{i-2},\tau_{i-1},c_i,c_{i+1})\biggr] \nonumber \\ 
&+& \sum_{\tau_{i+2},\tau_{i+3}} \biggl[P_4(c_i,\tau_{i+2},c_{i+1},\tau_{i+3})  
w(c_i,\tau_{i+2},c_{i+1},\tau_{i+3}) \nonumber \\
&-&P_4(c_i,c_{i+1},\tau_{i+2},\tau_{i+3}) 
w(c_i,c_{i+1},\tau_{i+2},\tau_{i+3})\biggr]
\label{eq-4}
\end{eqnarray}
governing the time evolution of the $2$-cluster probabilities 
$P_2(c_i,c_{i+1})$ involves the $4$-cluster probabilities 
$P_4(\tau_{i-1},\tau_i,\tau_{i+1},\tau_{i+2})$ for all those 
configurations which can lead to the $2$-cluster configuration 
$(c_i,c_{i+1};t)$ under consideration; 
\begin{eqnarray}
&w&\!\!\!\!(\tau_{1},\tau_{2},\tau_{3},\tau_{4}) = W(\tau_{1},\tau_{2},\tau_{3},\tau_{4} \rightarrow \tau_{1},\tau_{3},\tau_{2},\tau_{4}) \nonumber \\
&=& \min\biggl(1,e^{\beta J[(\tau_{1} \tau_{3}+\tau_{2} \tau_{4}) 
- (\tau_{1} \tau_{2}+\tau_{3} \tau_{4})+(\tau_{2}-\tau_{3})E]}\biggr) 
\label{eq-4a}
\end{eqnarray}
are the corresponding transition probabilities. However, the the 
Master equation governing the time evolution of the $4$-cluster 
probabilities $P_4(\tau_{i-1},\tau_i,\tau_{i+1},\tau_{i+2})$ 
involve $6$-cluster probabilities, and so on. A few concrete 
examples of such exact master equations for $n$-cluster 
probabilities are given in the appendix A. 
In the spirit of the cluster-mean-field approach, we truncate 
this hierarchy of exact Master equations by expressing, albeit 
approximately, all the $n+m$-cluster probabilities in terms of 
the $n$-cluster probabilities.

According to the definition of DH, the probability for a DH of $j$ 
is given by 
\begin{equation}
P(j) = P(\underline{1}|\underbrace{0000...0}_{j ~{\rm times}}1). 
\label{eq-dh1}
\end{equation}
We evaluate the right hand side of the equation (\ref{eq-dh1}) 
in the $2$-cluster and $4$-cluster approximations.

\subsection{Computer simulation}

In our computer simulations, we begin with initial conditions 
where the particles are arranged randomly on a lattice of 
linear size $L$ (i.e., the system is a linear chain of size $L$ 
in $d = 1$ and square lattice of size $L \times L$ in $d=2$). 
The system is then allowed to evolve, following the Kawasaki 
exchange algorithm with the Metropolis probabilities 
min$[1,e^{-\beta(\Delta {\cal H} + \ell E)}]$ mentioned above, 
for time steps of the order of $10^4$, so as to allow it to reach 
its steady-state and, then, for further 
$10^5$ steps during which the time-averaged flux is measured. 
Finally, for the same set of values of the parameters, this process 
is repeated for $100$ different initial configurations to compute 
the configuration-averaged flux. Since we did not observe any 
significant difference between the results for $L = 10^4$ and 
$L = 10^5$ in $d = 1$ and between those for $L=50$ and $L = 100$ 
in $d = 2$ all the data presented in this paper were generated 
using $L = 10^4$ in $d = 1$ and $L = 50$ in $d = 2$.

\section{1-cluster approximation in one-dimension} 

It has been realized for some time \cite{sz} that the smallest 
cluster one must consider in a dynamical cluster MFT depends on 
the nature of the dynamics. Since the Kawasaki dynamics conserves 
the number of particles, in principle, the smallest cluster must 
consist of at least one {\it pair} of sites. However, it is also 
known that for the $E \rightarrow \infty$ limit of the KLS model, 
which is exactly identical with the TASEP (with $q = 1$), the 
single-site MFT gives the exact result. Therefore, in this short 
section we not only establish explicitly the limitations of the 
$1$-cluster MFT at weak $E$ but also demonstrate how the accuracy 
of the $1$-cluster MFT increases with increasing $E$ in the KLS 
model in $d = 1$. 

The net flux is obtained from $F = F_f - F_r$ where the 
forward flux (i.e., flux in the direction of $\vec E$) is given by 
\begin{eqnarray}
F_f = \sum_{c_{i-1},c_{i+2}} P_4(c_{i-1},1,0,c_{i+2}) w(c_{i-1},1,0,c_{i+2}) 
\label{eq-5}
\end{eqnarray}
while the reverse flux (i.e., the flux against $\vec E$) is given by 
\begin{eqnarray}
F_r = \sum_{c_{i-1},c_{i+2}} P_4(c_{i-1},0,1,c_{i+2}) w(c_{i-1},0,1,c_{i+2}). 
\label{eq-6}
\end{eqnarray}

In the $1$-cluster approximation the $4$-cluster probabilities 
$P_4(c_{i-1},c_i,c_{i+1},c_{i+2})$ are approximated by the 
products of corresponding $1$-cluster probabilities. Therefore, 
utilizing the facts that $P_1(1) = c$ and $P_1(0) = 1-c$, in 
the $1$-cluster approximation, the equations (\ref{eq-5}) and 
(\ref{eq-6}) give, respectively, 
\begin{eqnarray}
F_f = \!\!\!\sum_{c_{i-1},c_{i+2}}\!\! P_1(c_{i-1}) c (1-c) P_1(c_{i+2}) w(c_{i-1},1,0,c_{i+2}) 
\label{eq-7}
\end{eqnarray}
and 
\begin{eqnarray}
F_r = \!\!\!\!\! \sum_{c_{i-1},c_{i+2}}\!\!\! P_1(c_{i-1}) (1-c) c P_1(c_{i+2}) w(c_{i-1},0,1,c_{i+2}) 
\label{eq-8}
\end{eqnarray}
for the forward and reverse flux. Hence, the net flux is
\begin{eqnarray}
F &=& c(1-c)\biggl[(2c^2-2c+1)(\min[1,e^{\beta E}] - \min[1,e^{-\beta E}]) \nonumber \\
&+& c(1-c)\biggl(\min[1,e^{\beta(E-4J)}] - \min[1,e^{\beta(-E+4J)}] \nonumber \\
&-& \min[1,e^{\beta(-E-4J)}] + \min[1,e^{\beta(E+4J)}]\biggr)\biggr].
\label{eq-9}
\end{eqnarray}
The expression (\ref{eq-9}) predicts that the net flux $F$ is 
independent of the sign of the interaction $J$, at all $c$ and 
$T$, irrespective of the strength of $E$; this is certainly not 
true in general, except at very large $E$ (fig.1). A comparison 
between the theoretical prediction (\ref{eq-9}) and the results  
of our computer simulations (fig.1) exposes not only the 
quantitative inaccuracy of the $1$-cluster MFT but also its 
failure to account for the qualitative features of $F(c)$ in 
the case of repulsive inter-particle interactions.

\begin{figure}[tb]
\includegraphics[width=\columnwidth]{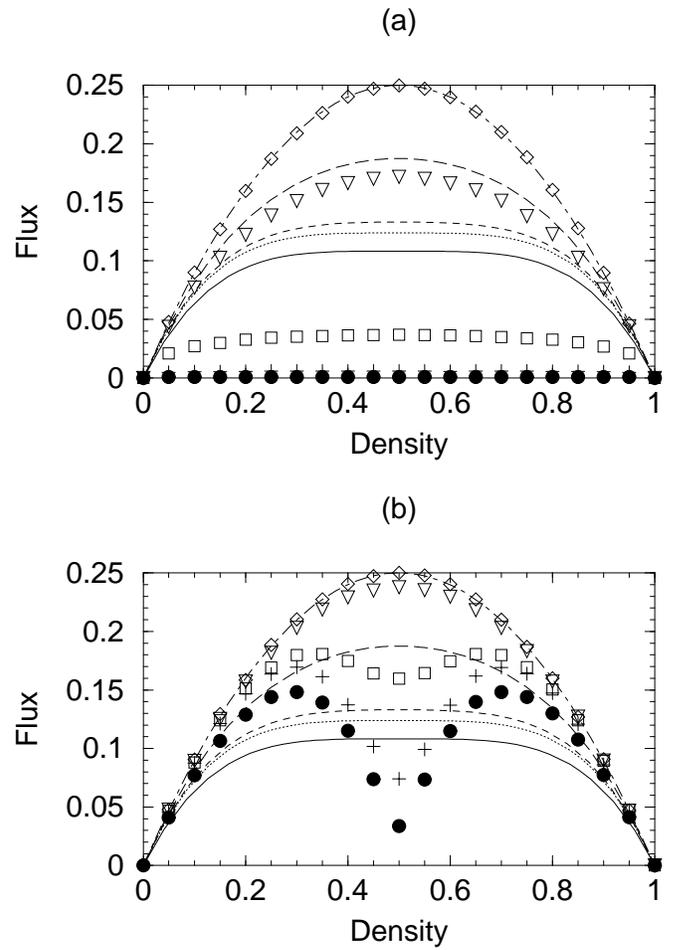}
\caption{Variation of the {\it net} flux $F$ with the density $c$ of 
the particles in the {\it steady-state} of the one-dimensional KLS 
model with (a) attractive (i.e., ferromagnetic) interaction $J = 1.0$ 
and (b) repulsive (i.e., anti-ferromagnetic) interaction $J = - 1.0$, 
both at $T = 0.5 |J|$. In both (a) and (b) the discrete data points 
have been obtained from our computer simulations with 
$E = 1.0 (\bullet), 2.0 (+) , 3.0 (\Box) , 4.0 ({\bigtriangledown})$ 
and $100.0 ({\Diamond})$, respectively. The lines represent 
the predictions of the $1$-cluster approximation for 
$E = 1.0$ (solid line), $2.0$ (dotted line), $3.0$ (dashed line), 
$4.0$ (long dashed) and $100.0$ (dot-dashed), respectively.}
\label{fig-1}
\end{figure}

The fact 
that the $1$-cluster MFT works better for stronger $E$ is not 
surprising as the one-dimensional KLS model reduces to the 
TASEP (with $q = 1$) in the limit $E \rightarrow \infty$ and 
it is well known that $1$-cluster MFT gives exact result for 
the TASEP. Thus, fig.\ref{fig-1} establishes the existence 
of strong correlations which are neglected by the $1$-cluster MFT.

\section{2-cluster approximation in one-dimension}

In the $2$-cluster approximation, we approximate the $4$-cluster 
probabilities, appearing in the exact Master equation for the 
2-cluster probabilities, by a product of $2$-cluster probabilities, 
i.e., 
\begin{equation}
P_4(\tau_{i-1},\tau_i,\tau_{i+1},\tau_{i+2}) \propto P_2(\tau_{i-1},\tau_i) P_2(\tau_i,\tau_{i+1}) P_2(\tau_{i+1},\tau_{i+2}),
\label{eq-10}
\end{equation}
or, more precisely, 
\begin{equation}
P_4(\tau_{i-1},\tau_i,\tau_{i+1},\tau_{i+2}) = P_2(\tau_{i-1}|\underline{\tau_i}) P_2(\tau_i,\tau_{i+1}) P_2(\underline{\tau_{i+1}}|\tau_{i+2}),
\label{eq-11}
\end{equation}
Next, we parametrize the $2$-cluster probabilities as follows:\\
\begin{equation}
P_2(0,0) = 1-c-a 
\label{eq-12}
\end{equation}
\begin{equation}
P_2(1,1) = c-a 
\label{eq-13}
\end{equation}
where
\begin{equation}
P_2(1,0) = P_2(0,1) = a; 
\label{eq-14}
\end{equation}
this parametrization enables us to satisfy the three constraints 
\cite{ssni} namely, $\sum_{c_{i+1}=0}^1 P_2(1,c_{i+1}) = P_1(1) = c$,
$\sum_{c_{i+1}=0}^1 P_2(0,c_{i+1}) = P_1(0) = 1-c$, and 
$P_2(1,0) = P_2(0,1)$. Equivalently, these constraints also imply 
that $dP_2(0,0)/dt = dP_2(1,1)/dt = - dP_2(0,1)/dt = - dP_2(1,0)/dt$.  
So, only one of the four equations represented by equation (\ref{eq-4}) 
can be taken as an independent equation. Thus, the  calculation of 
the four $2$-cluster probabilities boils down to the calculation 
of the single parameter $a$.

Using {\sl Mathematica\/} for an automated generation of the equations and 
simplification, we find that, in the steady state, $a$ satisfies 
the quadratic equation 
\begin{equation} 
A a^2 + B a + C = 0, 
\label{eq-15}
\end{equation} 
where 
\begin{eqnarray}
A &=& \min[1,e^{\beta(-E-4J)}] + \min[1,e^{\beta(E-4J)}] \nonumber \\
&-& \min[1,e^{\beta(-E+4J)}] - \min[1,e^{\beta(E+4J)}],
\label{eq-16}
\end{eqnarray}
\begin{equation}
B = - \biggl(\min[1,e^{\beta(-E-4J)}] + \min[1,e^{\beta(E-4J)}]\biggr), 
\label{eq-17}
\end{equation}
\begin{equation}
C = c(1-c)\{\min[1,e^{\beta(-E-4J)}] + \min[1,e^{\beta(E-4J)}]\}, 
\label{eq-18}
\end{equation} 
and, hence, taking the physical solution that allows the 
$2$-cluster probabilities to be between $0$ and $1$, we get 
\begin{equation}
a = (-B - \sqrt{B^2 - 4 A C})/(2A). 
\label{eq-19}
\end{equation}
The corresponding form of $P_2(1,0)$ for an alternative choice of 
$w$ was derived by Szabo et al.\cite{attila91}. Both the forms 
(\ref{eq-19}) and the equation (13) in reference \cite{attila91} 
are special cases of the general form \cite{attilacom}
\begin{equation}
a = (1 - \sqrt{1 - 2 c(1-c) \kappa})/\kappa 
\label{eq-19a}
\end{equation}
where 
\begin{equation}
\kappa = 2 \biggl[ 1 - \frac{w(\Delta H, -E) + w(\Delta H, +E)}{w(-\Delta H, -E)  + w(-\Delta H, +E)} \biggr];
\label{eq-19b}
\end{equation}
$w(\Delta H, \pm E)$ being the hopping probabilities against and 
along the field $E$, respectively.

Finally, in the $2$-cluster MFT, the forward flux and the reverse 
flux are given by 
\begin{eqnarray}
F_f = \sum_{c_{i-1},c_{i+2}} P_2(c_{i-1}|\underbar{1}) P_2(1,0) P_2(\underbar{0}|c_{i+2}) \nonumber \\
\times w(c_{i-1},1,0,c_{i+2}) 
\label{eq-20}
\end{eqnarray}
and
\begin{eqnarray}
F_r = \sum_{c_{i-1},c_{i+2}} P_2(c_{i-1}|\underbar{0}) P_2(0,1) P_2(\underbar{1}|c_{i+2}) \nonumber \\
\times w(c_{i-1},0,1,c_{i+2}), 
\label{eq-21}
\end{eqnarray}
respectively; the net flux is obtained from $F = F_f - F_r$. 

In the next few subsections we compare the predictions of this 
$2$-cluster MFT  with the corresponding numerical data obtained  
from our extensive computer simulations of the KLS model. 

\subsection{Fundamental diagrams for arbitrary E at $T > 0$} 

It is straightforward to verify that at an infinitely high 
temperature the expressions (\ref{eq-20}) and (\ref{eq-21}) for 
$F_f$ and $F_r$ become identical and, therefore, the net flux 
vanishes; hign temperature not only washes away the effects of 
$J$, as it is known to do even in equilibrium, but also the effects 
of $E$ making particle movement in both directions equally probable.

In fig.2 we plot the flux as a function of the particle density 
$c$, at a fixed non-zero finite temperature $T$, for five different 
values of $E$; the agreement between the theoretical prediction and 
computer simulation is very good for both attractive (fig.2a) as well 
as repulsive (fig.2b) interactions.

\begin{figure}[tb]
\includegraphics[width=\columnwidth]{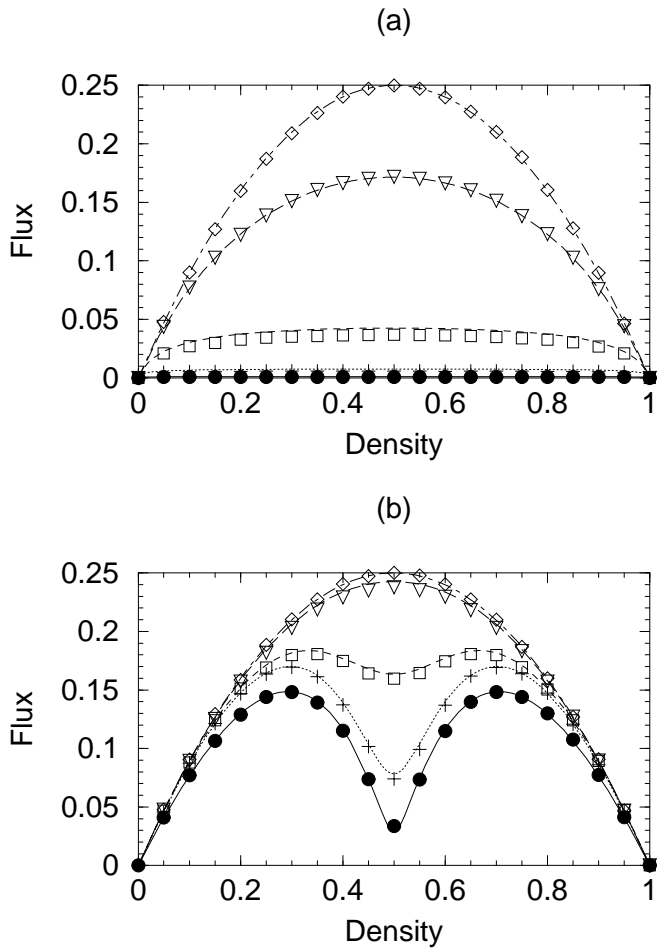}
\caption{Variation of the {\it net} flux $F$ with the density $c$ of 
the particles in the {\it steady-state} of the one-dimensional KLS 
model with (a) attractive interaction $J = 1.0$ and (b) repulsive 
interaction $J = - 1.0$, both at $T = 0.5 |J|$. In both (a) and (b) 
the discrete data points, have been obtained from our computer 
simulations with 
$E = 1.0 (\bullet), 2.0 (+) , 3.0 (\Box) , 4.0 ({\bigtriangledown})$ 
and $100.0 ({\Diamond})$, respectively. The lines represent 
the corresponding predictions of the $2$-cluster approximation.}
\label{fig-2}
\end{figure}

The $F(c)$ curves are the analogues of the fundamental relations 
for the traffic models. The shape of the curve $F(c)$ in the KLS 
model with {\it attractive} ($J > 0$) is qualitatively similar to 
those observed in the particle-hopping models of vehicular traffic. 
In sharp contrast, we find a qualitatively different shape of the 
curve $F(c)$ in the KLS model with {\it repulsive} interactions 
($J < 0$); there is a minimum, rather than maximum, at $c = 1/2$ 
provided the strength of $J$ is comparable to that of $E$. Moreover, 
for the same $T$ and $E$, the flux is higher in the case of 
{\it repulsive} inter-particle interactions than that in the case 
of {\it attractive} inter-particle interactions. Nevertheless, 
because of the particle-hole symmetry, the curves in both the 
figures \ref{fig-2}(a) and (b) are symmetric about $c = 1/2$ 
irrespective of the sign of the inter-particle interactions. 

Also note that in both the figures \ref{fig-2}(a) and (b) 
$F(c=1/2,E) \rightarrow 1/4$ as $E \rightarrow \infty$. 
This is a consequence of the fact that, as stated before, the KLS 
model reduces to the ASEP, with $q = 1$, in the limit 
$E \rightarrow \infty$ (in both the cases of attractive and 
repulsive inter-particle interactions) as long as $J$ and $T$ 
remain finite. In fact, in all the figures \ref{fig-1}(a), (b) 
and \ref{fig-2}(a), (b), the full curve $F(c,E\rightarrow \infty)$ 
is given by exact expression $F(c,E\rightarrow \infty) = c(1-c)$. 

Thus, unlike the $1$-cluster MFT, the $2$-cluster MFT 
reproduces the qualitative features of the $F(c)$ curves for all 
$E$ and for both attractive as well as repulsive interactions. 
Moreover, the flux predicted by the $2$-cluster MFT is also in 
good quantitative agreement with the corresponding computer 
simulation data, except for a narrow range of $c$ about $c=1/2$. 
This indicates that the predictions of $2$-cluster MFT, although 
quite accurate for arbitrary $E$, is {\it not exact} (except, of 
course, $E \rightarrow \infty$) for the KLS model in $d = 1$.
We shall improve our theory further by developing a $4$-cluster 
MFT in the next section. 

In order to get a deeper insight into the dependence of the flux 
$F(c,E,T)$ on $c,E,T$ as well as on the sign of $J$ we analyze 
the results of the $2$-cluster MFT in detail in a few special 
limits in the following subsections.

\subsection{Flux at T = 0}

Let us investigate the dependence of the flux $F(c,E,T=0)$ on the 
driving field $E$ at $T = 0$ for arbitrary values of $c$. For 
{\it repulsive} inter-particle interactions, the $2$-cluster 
expressions (\ref{eq-16})-(\ref{eq-18}) for $A, B$ and $C$ 
reduce to the simple forms $A = 2$, $B = -2$, $C = 2c(1-c)$ 
for all $E < 4|J|$, $A = 1$, $B = -2$, $C = 2c(1-c)$ at $E = 4|J|$ 
and to the forms $A = 0$, $B = -1$, $C = c(1-c)$ for all $E > 4|J|$. 
Substituting the zero-temperature values of $A, B$ and $C$ into 
the equation (\ref{eq-19}), we find that, at $T = 0$, for all 
$E < 4|J|$ $a \equiv P_2(1,0) = c$ for $c < 1/2$, 
and $a \equiv P_2(1,0) = 1-c$ for $c > 1/2$; 
physically, this means that when less than half of the sites are 
occupied by particles (holes), both the nearest-neighbors of 
each particle (hole) are {\it certainly} holes (particles) 
because of the repulsive nature of the inter-particle interaction. 
On the other hand, at $T = 0$, for all $E > 4|J|$, 
$a \equiv P_2(1,0) = c(1-c)$ for any arbitrary $c$; this implies 
that at $T = 0$ the effects of all $E > 4|J|$ is equivalent to 
those of infinitely large $E$ at all finite non-zero $T$ so that 
$1$-cluster MFT becomes exact. Moreover, at $T = 0$, for $E = 4|J|$, 
$a \equiv P_2(1,0) = 1 - \sqrt{1 - 2c(1-c)}$ for all $c$.

Substituting the appropriate expression of $a$, derived above for 
$T = 0$, into the limiting form of the net flux $F(c,E,T=0)$, 
obtained from the $2$-cluster expressions (\ref{eq-20}) and 
(\ref{eq-21}), we get 
\begin{eqnarray}
F(c,E,T=0) = \left\{ \begin{array}{ll} 
~F^{r}_{<}(c) & \mbox{for $E < 4|J|$,}\\ 
~F^{r}_{=}(c) & \mbox{for $E = 4|J|$,}\\ 
~F^{r}_{>}(c) & \mbox{for $E > 4|J|$,} 
             \end{array}
                \right.
\label{eq-22}
\end{eqnarray}
for the KLS model with {\it repulsive} inter-particle interactions 
at $T = 0$, where 
\begin{eqnarray}
F^{r}_<(c) = \theta(0.5-c) ~\biggl[\frac{c}{1-c}(1-2c)\biggr] \nonumber \\ 
+ \theta(c-0.5) ~\biggl[\frac{1-c}{c}(2c-1)\biggr], 
\label{eq-23}
\end{eqnarray}
\begin{equation}
F^{r}_{=}(c) = \frac{2\{2c(1-c)-1\}+2\sqrt{1-2c(1-c)}\{1-c(1-c)\}}{c(1-c)},
\label{eq-23b}
\end{equation}
and 
\begin{equation}
F^{r}_>(c) = c(1-c);
\label{eq-24}
\end{equation}
$\theta(z)$ being the step function, namely, $\theta(x) = 0$ for 
all $x < 0$ and $\theta(z) = 1$ for all $x > 0$.

In the special case of half-filling, the particles remain ``pinned" 
to their respective positions by the inter-particle interactions 
$J$ and no $E < 4|J|$ is strong enough to cause any ``de-pinning". 
Such ``switching" of the flux from zero to a non-zero value 
{\it discontinuously}, in response to the driving field $E$, 
has been reported earlier for the KLS model on a linear chain 
\cite{attila91} as well as on a square lattice \cite{dickman}.

In sharp contrast to the $c$-dependence of $F^{r}_<(c)$ in the case 
of {\it repulsive} inter-particle interactions, we have 
$F^{a}_<(c) = 0$ for all $c$ when the inter-particle interaction is 
{\it attractive}; therefore, in the latter case, 
\begin{eqnarray}
F(c,E,T=0) = \left\{ \begin{array}{ll} 
~F^{a}_{<}(c) & \mbox{for $E < 4|J|$,}\\ 
~F^{a}_{=}(c) & \mbox{for $E = 4|J|$,}\\ 
~F^{a}_{>}(c) & \mbox{for $E > 4|J|$,} 
             \end{array}
                \right.
\label{eq-25}
\end{eqnarray}
for the KLS model with {\it attractive} inter-particle interactions 
at $T = 0$, where 
\begin{eqnarray}
F^{a}_<(c) = 0, 
\label{eq-26a}
\end{eqnarray}
\begin{equation}
F^{a}_{=}(c) = \frac{\{2c(1-c)+1\}-\sqrt{1+4c(1-c)}}{2c(1-c)},
\label{eq-26b}
\end{equation}
and 
\begin{equation}
F^{a}_>(c) = c(1-c).
\label{eq-26}
\end{equation}
At $T = 0$, all $E > 4|J|$ is equivalent to $E \rightarrow \infty$, 
irrespective of the sign of $J$ and, hence, the corresponding flux 
is $c(1-c)$.

\subsection{Temperature-dependence of flux}

\begin{figure}[tb]
\includegraphics[width=\columnwidth]{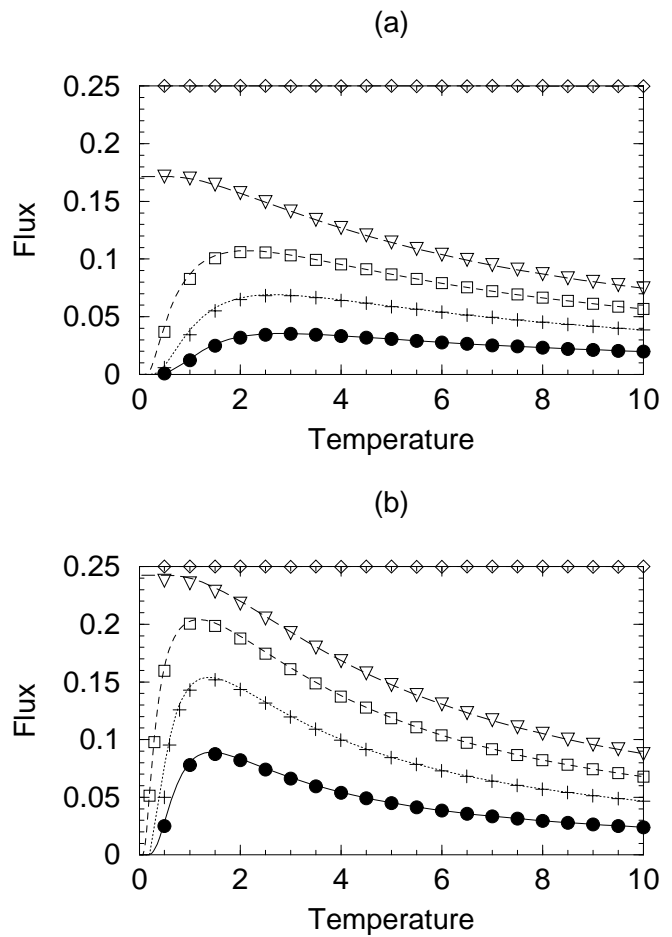}
\caption{Variation of the {\it net} flux $F$ with the temperature 
$T$ in the {\it steady-state} of the one-dimensional KLS model 
with (a) attractive interaction $J = 1.0$ and (b) repulsive 
interaction $J = - 1.0$, both at $c = 0.5$. In both (a) and (b) the 
discrete data points, have been obtained from our computer simulations 
with $E = 1.0 (\bullet), 2.0 (+) , 3.0 (\Box) , 4.0 ({\bigtriangledown})$ 
and $100.0 ({\Diamond})$, respectively. The lines represent 
the corresponding predictions of the $2$-cluster approximation.}
\label{fig-3}
\end{figure}

\begin{figure}[tb]
\includegraphics[width=\columnwidth]{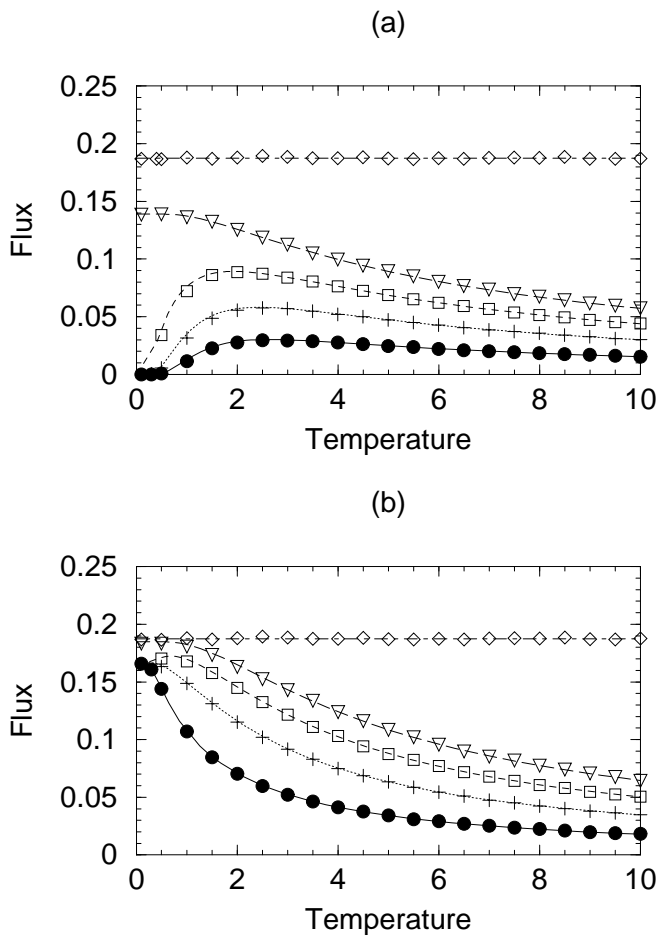}
\caption{Same as in fig.(\ref{fig-3}), except that the density 
of the particles is $c = 0.25$. }
\label{fig-4}
\end{figure}

In this subsection we consider the dependence of $F$ on $T$ at a 
few special values of $c$. As demonstrated in the figures 
(\ref{fig-3})-(\ref{fig-5}), the predictions of the $2$-cluster 
theory agree very well with the corresponding numerical data 
obtained from our computer simulations.

The non-monotonic variation of the flux with temperature at small 
and intermediate values of $E$ is an interesting phenomenon. So 
long as $T$ is non-zero but much smaller than $E$, it works against 
$J$ and helps in ``de-pinning" the particles which can move forward 
under the influence of the driving field $E$. However, when $T$ 
becomes much larger than $E$, then it washes out the effect of $E$ 
allowing particles to move against $\vec E$ as often as along $\vec E$. 

\subsection{DH distribution} 

In the $1$-cluster approximation, the DH distribution is given by 
\begin{equation}
P_{1c}(j) = c(1-c)^{j}.
\label{eq-dh2}
\end{equation}
However, in the $2$-cluster approximation, we write 
\begin{equation}
P_{2c}(j) = P_2(\underline{1}|1), ~{\rm for} ~j = 0 
\label{eq-dh3}
\end{equation}
and 
\begin{equation}
P_{2c}(j) = P_2(\underline{1}|0)\{P_2(\underline{0}|0)\}^{j-1} P_2(\underline{0}|1), ~{\rm for} ~j \geq 1. 
\label{eq-dh4}
\end{equation}
Hence, in the $2$-cluster approximation, we get
\begin{equation}
P_{2c}(j) = 1 - \frac{a}{c}, ~{\rm for} ~j = 0 
\label{eq-dh5}
\end{equation}
whereas 
\begin{equation}
P_{2c}(j) = \frac{a^2}{c(1-c)} \biggl[1 - \frac{a}{1-c}\biggr]^{j-1}, ~{\rm for} ~j \geq 1.
\label{eq-dh6}
\end{equation}

\begin{figure}[tb]
\includegraphics[width=\columnwidth]{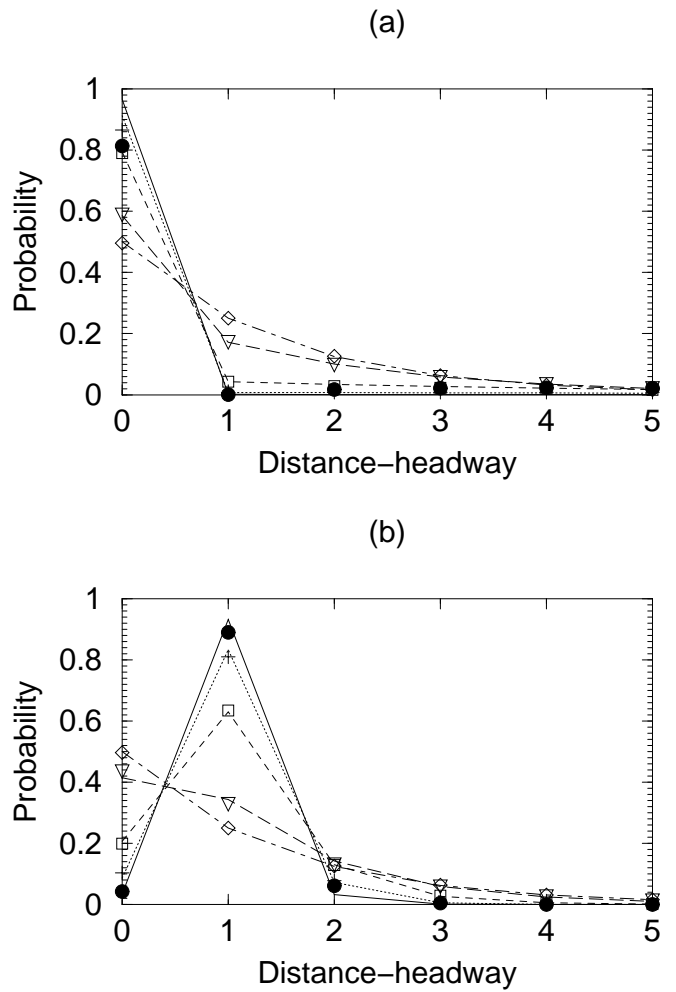}
\caption{The distance-headway distribution in the {\it steady-state} 
of the one-dimensional KLS model with (a) attractive interaction 
$J = 1.0$ and (b) repulsive interaction $J = - 1.0$, both at 
for $c = 1/2$ and $T = 0.5 |J|$. In both (a) and (b) the discrete 
data points have been obtained from our computer simulations with 
$E = 1.0 (\bullet), 2.0 (+) , 3.0 (\Box) , 4.0 ({\bigtriangledown})$ 
and $100.0 ({\Diamond})$, respectively. The lines represent 
the corresponding predictions of the $2$-cluster approximation.}
\label{dhfig1}
\end{figure}

\begin{figure}[tb]
\includegraphics[width=\columnwidth]{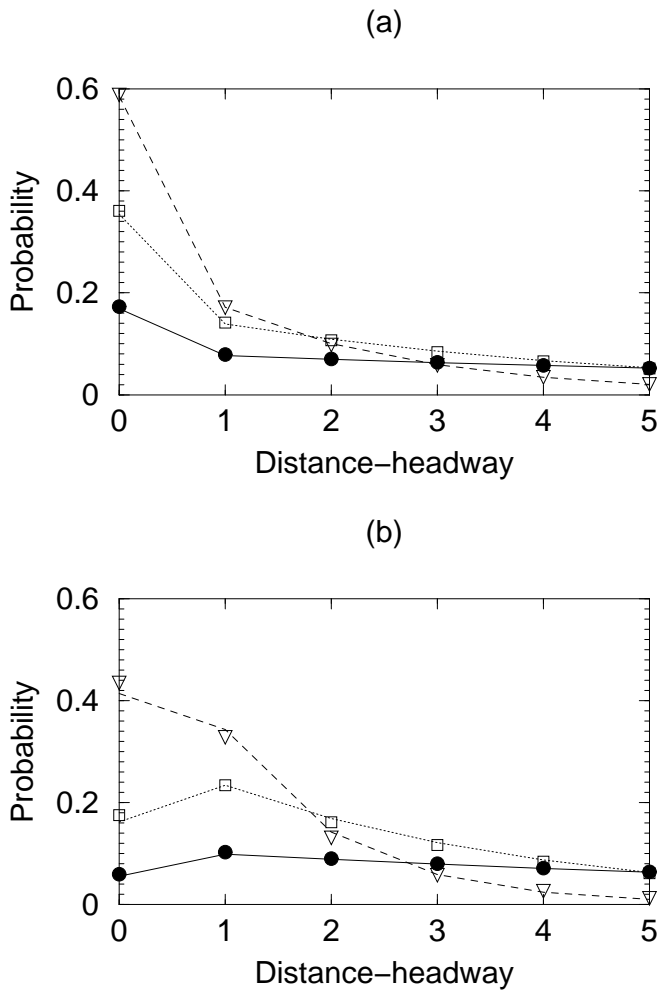}
\caption{The distance-headway distribution in the {\it steady-state} 
of the one-dimensional KLS model with (a) attractive interaction 
$J = 1.0$ and (b) repulsive interaction $J = - 1.0$, both at 
for $E = 4.0$ and $T = 0.5 |J|$. In both (a) and (b) the discrete 
data points have been obtained from our computer simulations with 
$c = 0.1 (\bullet), 0.25 (\Box)$ and $0.5 ({\bigtriangledown})$, 
respectively. The lines represent the corresponding predictions of 
the $2$-cluster approximation.}
\label{dhfig2}
\end{figure}

The variations of the DH distribution with $E$ (for fixed $c$) and 
with $c$ (for fixed $E$), as predicted by (\ref{eq-dh5}) and 
(\ref{eq-dh6}) of the $2$-cluster approximation, are compared with 
the corresponding computer simulation data in fig.\ref{dhfig1}.
and fig.\ref{dhfig2}, respectively. Note that for $c = 0.5$, in the 
absence of $E$, the most probable DH is $j = 0$ or $j = 1$ 
depending on whether the interaction is attractive (i.e., 
ferromagnetic, in the language of magnetism) or repulsive (i.e., 
antiferromagnetic). This type of spatial organization of the 
particles persists even in the presence of $E$ as long as $E$ 
is much weaker than the strength of the interaction $J$ 
(see fig.\ref{dhfig1}). However, deviation from this spatial 
organization increases gradually with increasing strength of 
$E$. In the limit $E \rightarrow \infty$, for all finite $|J|$, 
the DH distribution approaches the exact DH distribution of the 
TASEP and, as expected, is independent of the sign of the 
interaction $J$. For a given $E$, which is comparable with the 
strength $|J|$ of the interaction, increasing $c$ leads to more 
congestion and, therefore, the probability of having a DH = 0 
becomes larger for higher $c$ (se fig.\ref{dhfig2}). 

\section{4-cluster approximation in one-dimension}

The exact master equation for the $4$-cluster probabilities  
(given in the appendix B), as expected on general grounds, 
involve $6$-cluster probabilities. In the $4$-cluster 
approximation we break up the $6$-cluster probabilities in 
terms of the products of the $4$-cluster probabilities using 
the prescription 
\begin{eqnarray}
P_6(\tau_{i-2},\tau_{i-1},\tau_i,\tau_{i+1},\tau_{i+2},\tau_{i+3})& = & P_4(\tau_{i-2},\tau_{i-1}|\underline{\tau_i,\tau_{i+1}}) \nonumber \\
   \times  P_4(\tau_{i-1},\tau_i,\tau_{i+1},\tau_{i+2}) && \nonumber \\
   \times  P_4(\underline{\tau_{i},\tau_{i+1}}|\tau_{i+2},\tau_{i+3}). &&
\label{eq-27}
\end{eqnarray}

\begin{figure}[tb]
\includegraphics[width=\columnwidth]{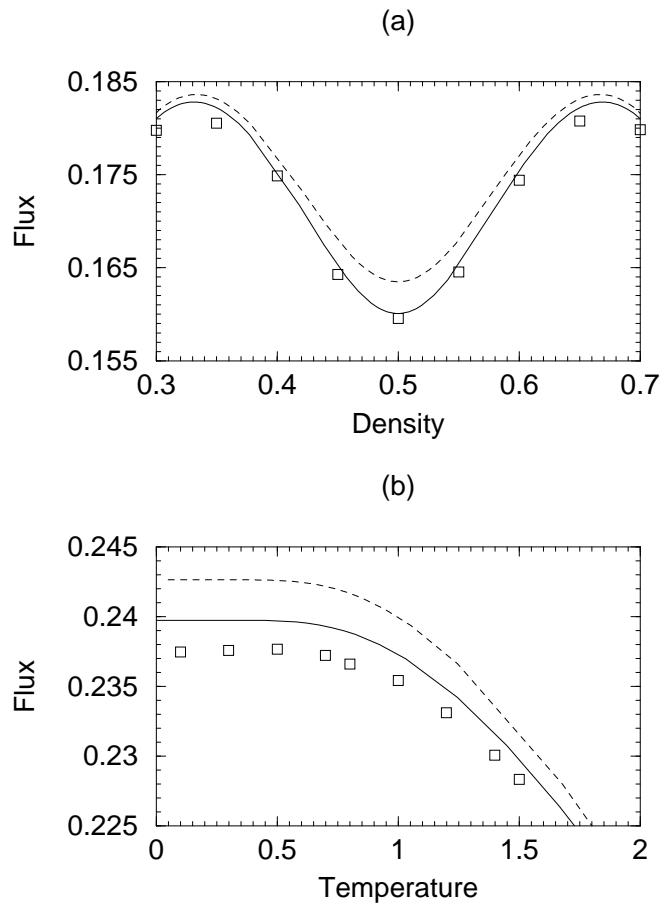}
\caption{Comparison of the predictions of $2$-cluster and $4$-cluster 
MFTs on the variation of the {\it net} flux $F$ with (a) the density 
and (b) temperature in the {\it steady-state} of the one-dimensional 
KLS model with repulsive interaction $J = - 1.0$. The parameters in 
(a) are $T = 0.5, E = 3.0$ while those in (b) are $c = 0.5, E = 4.0.$ 
The dashed and solid lines are the theoretical results obtained in 
the $2$-cluster and $4$-cluster approximations, respectively, whereas 
the discrete data points are the numerical data obtained from our 
computer simulations.}
\label{fig-5}
\end{figure}

Due to the constraints of conditional probabilities and
conservation of total probabilities, at $4$-cluster level, we
have 7 independent equations and the same number of
variables.  The solution is obtained by solving a set of
non-linear equations.
We find that although the $4$-cluster results are, clearly, an 
improvement over the $2$-cluster results, the difference in the 
actual numerical values of the flux in the two approximations, 
for the same set of parameters, is extremely small. The flux 
obtained in these two approximations for a typical set of 
values of the parameters are compared in fig.\ref{fig-5} (the 
corresponding differences in the DH distributions are too 
small to be shown in a figure).

It is worth pointing out that the $4$-cluster approximation is 
not only a {\it quantitative} improvement over the $2$-cluster 
approximation. The $4$-cluster approximation can account for 
the forward-backward symmetry breaking, an interesting phenomenon 
\cite{attila91,attila93}, which the $2$-cluster approximation 
fails to capture. 

\section{Cluster approximations in two dimension}

It is well known \cite{sz} that, usually, the cluster MFT fails   
to account for the properties of the driven-diffusive lattice 
gases below the ordering temperature $T_c(E)$. Therefore, we 
confine our discussions in this section to temperatures 
$T > T_c(E)$. 

\begin{figure}[tb]
\includegraphics[width=\columnwidth]{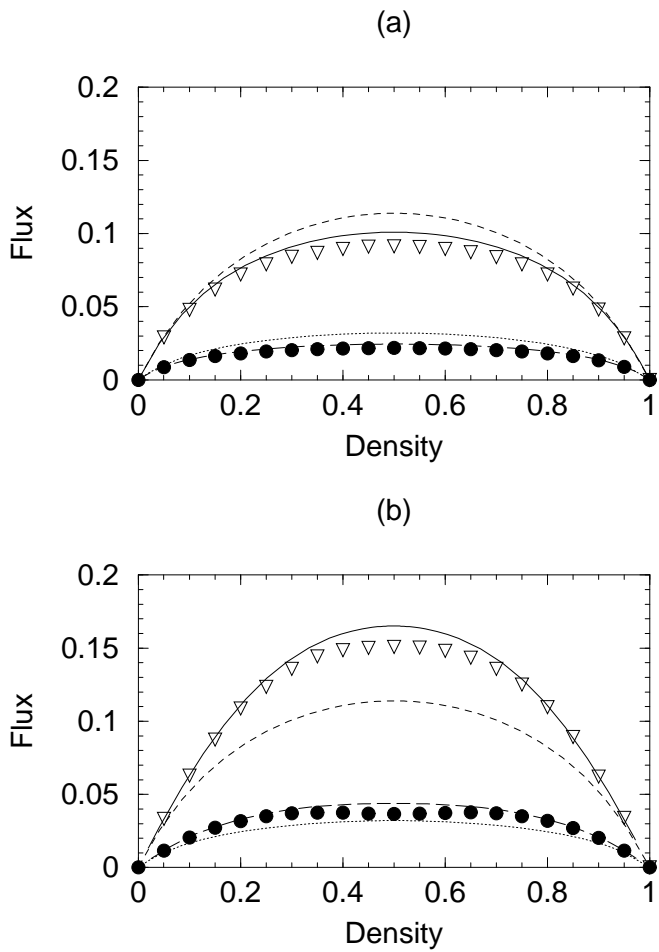}
\caption{Variation of the {\it net} flux $F$ with the density $c$ of 
the particles in the {\it steady-state} of the two-dimensional KLS 
model with (a) attractive interaction $J = 1.0$ and (b) repulsive 
interaction $J = - 1.0$, both at $T = 1.5 T_c(E=0)$. In both (a) and 
(b) the discrete data points have been obtained from our computer 
simulations with $E = 1.0 (\bullet)$ and $4.0 ({\bigtriangledown})$, 
respectively. In both (a) and (b) the dotted and dashed lines represent 
the predictions of the $1$-cluster MFT for $E = 1.0$ and $4.0$, 
respectively, while the long-dashed and solid lines represent the 
corresponding predictions of the $2$-cluster MFT.}
\label{fig-6}
\end{figure}

As in $d = 1$, for the same set of parameters, the flux in the 
two-dimensional KLS model with attractive inter-particle interactions 
(fig.\ref{fig-6}(a)) is lower than that in the same model with 
repulsive inter-particle interactions (fig.\ref{fig-6}(b)).

\begin{figure}[tb]
\includegraphics[width=\columnwidth]{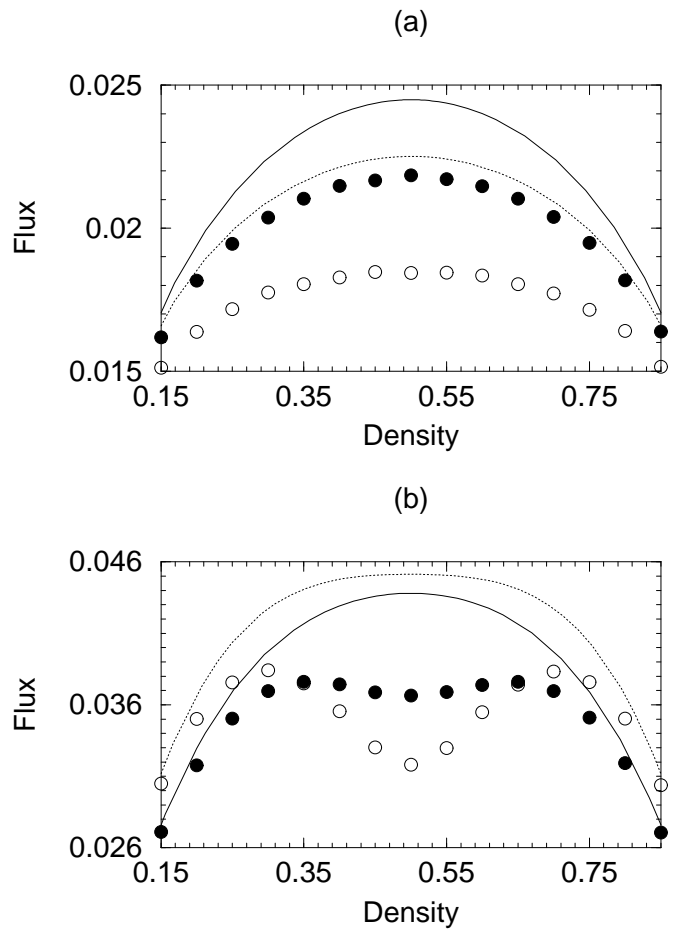}
\caption{Comparison of the predictions of $1$-cluster and $2$-cluster 
MFTs on the {\it net} flux $F$ in the {\it steady-state} of the 
two-dimensional KLS model with (a) attractive interaction $J = 1.0$ 
and (b) repulsive interaction $J = 1.0$, all for $E = 1.0$. In both 
(a) and (b) the discrete data points have been obtained from our 
computer simulations at $T = 1.25 T_c(E=0) (\circ)$ and 
$T = 1.5 T_c(E=0) (\bullet)$, respectively. In both (a) and (b) the 
dotted and solid lines represent the predictions of the $2$-cluster 
MFT at $T = 1.25 T_c(E=0)$ and $T = 1.5 T_c(E=0)$, respectively.}
\label{fig-7}
\end{figure}

A comparison of the predictions of the cluster MFT 
with the computer simulation data (fig.\ref{fig-6}) establishes 
that, in the case of {\it attractive} interactions both the 
$1$-cluster and $2$-cluster MFT over-estimate the flux, although 
the prediction of the $2$-cluster theory is closer to the computer 
simulation data. On the other hand, in the case of {\it repulsive} 
interactions, the $1$-cluster MFT gives an under-estimate whereas 
the $2$-cluster MFT provides an over-estimate of the flux. The 
level of accuracy of the $2$-cluster MFT can be estimated from 
the plots in fig.(\ref{fig-7}); a close inspection, thus, reveals 
that the $2$-cluster MFT does not reproduce the dip in the flux 
around $c = 1/2$ in the case of repulsive inter-particle interactions.

\section{Comparison with the results for other models}

It is well established \cite{derrida} that the $1$-cluster MFT 
result (Eq.\ref{eq-3}) is the {\it exact} expression for the 
flux in TASEP. If the random-sequential updating of the TASEP 
is replaced by the parallel updating it becomes identical with 
the Nagel-Schreckenberg (NS) model \cite{ns} of vehicular traffic 
with $V_{max} = 1$, where $V_{max}$ is the largest (integer) speed 
allowed for each of the vehicles.

In the case of the NS model 
the $1$-cluster MFT makes an under-estimate of the flux but 
the $2$-cluster MFT treatment is adequate to take into account 
the correlations introduced purely by the parallel dynamics and, 
therefore, it gives the exact result \cite{ssni}. 

The KLS model can be regarded as an extension of the TASEP by 
incorporating non-vanishing 
inter-particles interactions through non-zero $J$. Our 
results reported in this paper show that neither the 
$1$-cluster MFT nor the $2$-cluster MFT yield exact flux in 
the KLS model; although the $2$-cluster results are accurate 
to order $10^{-3}$ it is the $4$-cluster results that are 
practically indistinguishable from the corresponding computer 
simulation data. Although it is possible that the results of 
the $4$-cluster MFT might be exact for the KLS model we refrain 
from making such a claim as we do not have any rigorous proof.

Note that in the special case 
\begin{eqnarray}
E &=& 0 \nonumber \\
\beta J &=& (1/4) \ln ~p \quad (\rm{with\ } 0 \leq p < 1), 
\label{eq-28}
\end{eqnarray}
we have $A = 2 q$, $B = -2$ and $C = 2 c (1-c)$ where $q = 1-p$.  
In this case, the quadratic equation (\ref{eq-16}) reduces to 
the form $q a^2 - y + c(1-c) = 0$ which was derived \cite{ssni} 
directly from the $2$-cluster MF treatment of the NS model with 
$V_{max} = 1$.  Interestingly, the $2$-cluster result is not exact 
for the KLS model, but it gives exact result for the NS model. 
This is a consequence of the fact that the $2$-cluster MFT gives, 
in general exact results for the one-dimensional Ising model in 
equilibrium. 

\begin{figure}[tb]
\includegraphics[width=\columnwidth]{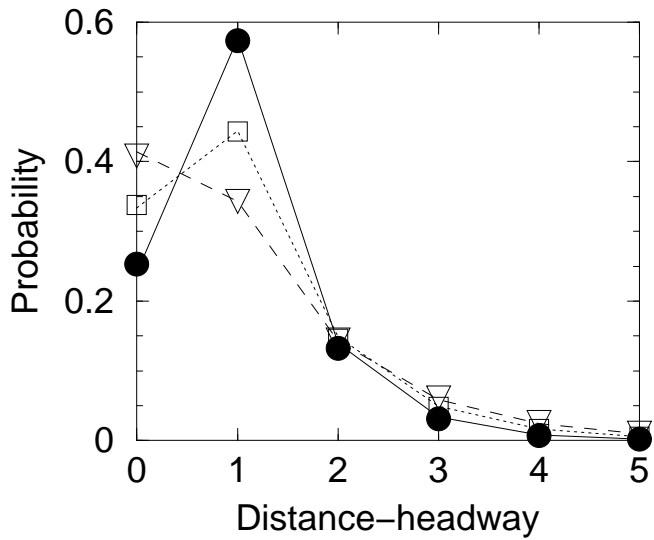}
\caption[Headway]{The discrete data points represent the distance-headway 
distribution in the steady-state of the one-dimensional 
KLS model with repulsive interaction $J = - 1.0$, for $c = 1/2$ 
at three different values of $T = 4 J/(\ln p)$ corresponding, 
respectively, to $p = 0.1$ ($\bullet$), $p = 0.25$ ($\Box$) and 
$p = 0.5$ ($\bigtriangledown$), all for $E = 0.0$. The lines are 
merely guides to the eye connecting the points corresponding to 
the DH distributions in the NS model with $V_{max} = 1$ and 
$p = 0.1$ (solid line), $p = 0.25$ (dotted line) and $p = 0.5$ 
(dashed line), respectively. All the results in this figure have 
been obtained in the $2$-cluster approximation.}
\label{figns}
\end{figure}

Since $p < 1$, the relation (\ref{eq-28}) maps the NS model, 
with $V_{max} = 1$, onto an {\it antiferro}-magnetic Ising model, 
(or, equivalently, to the KLS model, with {\it repulsive} 
inter-particle interactions, in the absence of external drive) 
so that the steady-state of the former is identical to the 
equilibrium state of the latter. Consequently, for all densities 
$c$, we observe perfect agreement of the DH distributions in 
the NS model and that in the KLS model with $E = 0$, 
$\beta J = (1/4) \ln ~p$ (fig.\ref{figns}).

Since the relation (\ref{eq-28}) maps the NS model (with 
$V_{max} = 1$) onto the KLS model only for $E = 0$, this 
mapping cannot relate the properties of the KLS model for 
any non-zero $E$ with those of the NS model. Interestingly, 
in the NS model with $V_{max} = 1$ the flux is {\it maximum} 
at $c = 1/2$ for all $q$. In sharp contrast, the flux is {\it 
minimum} at $c = 1/2$ in the KLS model with repulsive  
(anti-ferromagnetic) interactions so long as $E$ is not much 
stronger than $|J|$; as $E$ increases the depth of the well 
at $c = 1/2$ in fig.\ref{fig-2} decreases and, eventually, for 
sufficiently large $E$ the flux exhibits its maximum at $c = 1/2$. 
Moreover, the location of the maximum in the DH distribution 
depends crucially on the sign of the interaction $J$ provided 
$E$ is not much larger than $|J|$. 

The time-interval between the arrivals (or departures) of the 
successive particles at a detector site is defined to be the 
corresponding time-headway {\bf TH}. The exact TH distribution 
for the NS model with $V_{max} = 1$ has been calculated 
\cite{chowth}. However, because of the possibility of hopping 
of the particles against $E$ in the KLS model, the analytical 
calculation of TH distribution is extremely difficult and will 
not be reported here.

\section{Conclusion}

In this paper we have reported the results of cluster-mean-field 
theoretic treatments of a driven-diffusive lattice gas model to 
calculate the flux of the particles under the influence of the 
driving field. Although we have considered the standard model, 
namely, the Katz-Lebowitz-Spohn model in this paper, our technique 
is sufficiently general that it can be used, in principle, to 
calculate the flow properties of any other driven-diffusive 
lattice gas. 

We have shown interesting novel features of the flux-density 
relation of the KLS model with {\it repulsive} inter-particle 
interactions. Our investigation has helped in elucidating the 
roles of inter-particle interactions $J$, temperature $T$ and 
the driving field $E$ in determining the trend of variation of 
the flux with the density of the particles in the driven-diffusive 
lattice gas models.

\section*{Acknowledgment} One of us (DC) would like to thank the 
Department of Computational Science of the National University 
of Singapore for hospitality during a short visit  
where this paper was completed. We also thank A. Schadschneider 
and A. Szolnoki for useful correspondence.

\section{Appendix A}

Concrete examples of exact master equations for $n$-cluster 
probabilities in the one-dimensional KLS model:\\

\begin{eqnarray}
\frac{dP_2(0,0)}{dt} &=& -\min[1,e^{-\beta(E+4J)}] P_4(0,0,1,1) \nonumber \\
&+& \min[1,e^{\beta(E+4J)}] P_4(0,1,0,1) \nonumber \\
&+& \min[1,e^{\beta(-E+4J)}] P_4(1,0,1,0) \nonumber \\
&-& \min[1,e^{\beta(E-4J)}] P_4(1,1,0,0),
\label{eq-29} 
\end{eqnarray}

\begin{eqnarray}
\frac{dP_3(0,0,0)}{dt} &=& -\min[1,e^{-\beta E}] P_5(0,0,0,1,0) \nonumber \\
&-&\min[1,e^{-\beta(E+4J)}] P_5(0,0,0,1,1) \nonumber \\
&+&\min[1,e^{-\beta E}] P_5(0,0,1,0,0) \nonumber \\
&+&\min[1,e^{\beta E}] P_5(0,0,1,0,0) \nonumber \\
&+& \min[1,e^{\beta(E+4J)}] P_5(0,0,1,0,1) \nonumber \\
&-&\min[1,e^{\beta E}] P_5(0,1,0,0,0) \nonumber \\
&+& \min[1,e^{\beta(-E+4J)}] P_5(1,0,1,0,0) \nonumber \\
&-& \min[1,e^{\beta(E-4J)}] P_5(1,1,0,0,0).
\label{eq-30} 
\end{eqnarray}

\section{Appendix B} 

Exact master equations for the $4$-cluster probabilities in 
the one-dimensional KLS model: \\

\begin{eqnarray}
&\biggl[&\frac{dP_4(c_{i-1},c_i,c_{i+1},c_{i+2})}{dt}\biggr] \nonumber \\
&=& \sum_{\tau_{i-2},\tau_{i+3}} \biggl[P_6(\tau_{i-2},c_{i-1},c_{i+1},c_i,c_{i+2},\tau_{i+3}) \nonumber \\ 
&\times& w(c_{i-1},c_{i+1},c_i,c_{i+2}) \nonumber \\ 
&-&  P_6(\tau_{i-2},c_{i-1},c_i,c_{i+1},c_{i+2},\tau_{i+3}) \nonumber \\ 
&\times& w(c_{i-1},c_i,c_{i+1},c_{i+2})\biggr] \nonumber \\ 
&+& \sum_{\tau_{i-2},\tau_{i+3}} \biggl[P_6(\tau_{i-2},c_{i-1},c_i,c_{i+2},c_{i+1},\tau_{i+3}) \nonumber \\ 
&\times& w(c_{i-1},c_i,c_{i+2},c_{i+1}) \nonumber \\ 
&-& P_6(\tau_{i-2},c_{i-1},c_i,c_{i+1},c_{i+2},\tau_{i+3}) \nonumber \\ 
&\times& w(c_{i-1},c_i,c_{i+1},c_{i+2})\biggr] \nonumber \\ 
&+& \sum_{\tau_{i-2},\tau_{i+3}} \biggl[P_6(\tau_{i-2},c_i,c_{i-1},c_{i+1},c_{i+2},\tau_{i+3}) \nonumber \\ 
&\times& w(c_i,c_{i-1},c_{i+1},c_{i+2}) \nonumber \\ 
&-& P_6(\tau_{i-2},c_{i-1},c_i,c_{i+1},c_{i+2},\tau_{i+3}) \nonumber \\ 
&\times& w(c_{i-1},c_i,c_{i+1},c_{i+2})\biggr] \nonumber \\
&+& \sum_{\tau_{i+3},\tau_{i+4}} \biggl[P_6(c_{i-1},c_i,c_{i+1},\tau_{i+3},c_{i+2},\tau_{i+4}) \nonumber \\ 
&\times& w(c_{i+1},\tau_{i+3},c_{i+2},\tau_{i+4}) \nonumber \\ 
&-& P_6(c_{i-1},c_i,c_{i+1},c_{i+2},\tau_{i+3},\tau_{i+4}) \nonumber \\ 
&\times& w(c_{i+1},c_{i+2},\tau_{i+3},\tau_{i+4})\biggr] \nonumber \\
&+& \sum_{\tau_{i-2},\tau_{i-3}} \biggl[P_6(\tau_{i-3},c_{i-1},\tau_{i-2},c_i,c_{i+1},c_{i+2}) \nonumber \\ 
&\times& w(\tau_{i-3},c_{i-1},\tau_{i-2},c_i)\nonumber \\  
&-& P_6(\tau_{i-3},\tau_{i-2},c_{i-1},c_i,c_{i+1},c_{i+2}) \nonumber \\ 
&\times& w(\tau_{i-3},\tau_{i-2},c_{i-1},c_i)\biggr]. 
\label{eq-31}
\end{eqnarray}

\bibliographystyle{plain}

\end{document}